# IMPLEMENTATION OF JFPC (JAVA FOR PROCESS CONTROL) UNDER LINUX – A MECHANISM TO ACCESS INDUSTRIAL I/Os IN A JAVA/LINUX-ENVIRONMENT

H. Kleines, P. Wüstner, K. Settke, K. Zwoll,
Forschungszemtrum Jülich, D-52425 Jülich, Germany

Abstract

Using JAVA for process control requires the access to I/O-Hardware. This problem is addressed in several ongoing standardisation efforts for Real Time Java. A spin-off of these standardisation efforts is the Siemens product JFPC (Java for Process Control), which is based on the ideas of OPC (OLE for Process Control). It is available under the proprietary RT kernel RMOS and under Windows NT. Based on a contract with Siemens, ZEL - the central electronics facility of Forschungszentrum Juelich – has implemented JFPC under Linux. Additionally, several JFPC providers, e.g for accessing industrial components via PROFIBUS DP have been implemented.

## 1 INTRODUCTION

The Java programming language has become popular also in control applications for physics experiments, because of its simplified object model, its safety and its relative platform independence [1]. Forschungszentrum Jülich uses Java for small control applications, too. Because of Java`s security concepts and the abstraction from the underlying system, explicit access to I/O hardware is not supported in a pure Java environment. The product JFPC (Java for Process Control) which has been developed by Siemens for Windows NT and their proprietary real time kernel RMOS, offers an OPC-like access to process data. Forschungszentrum Jülich has done a port of major parts of JFPC to Linux in order to access industrial process periphery from an Linux-based Java application

## 2 HARDWARE ACCESS VIA JAVA

Explicit hardware access is not supported by the extensive Java APIs, but it is required in typical embedded and RT (real time) applications. So several hardware access APIs have been defined in the ongoing specification work for RT Java.

Besides the Java standardization in the "Java Community Process", which is controlled by SUN, there is standardization work going on in the J-Consortium, an independent organization of companies (HP, Microsoft, Siemens, Aonix, Omron,.....). The primary goal of the J-Consortium is to promote the development of standards for embedded and RT Java technologies. The J-Consortium was founded to get independent of SUNs intellectual property rights and to achieve a higher degree of openness and vendor-neutrality.

The outcome of the Java Community Process was a draft for the "RT Specification for Java" (RTSJ) [2], published by the "RT for Java Expert Group" as JSR-000001. RTSJ also defines I/O-access according to a simple flat approach: Typed data can be read/written by method calls of a RawPhysicalMemory object, by specifying the offset in the corresponding memory area.

The RT Java working group of the J-Consortium produced the standards draft covering the "RT Core" extensions to Java [3], which provide RT programming capabilities to the Java platform, as RTSJ does. But both documents are incompatible.

Because almost all applications in the domain of industrial automation require some kind of I/O access, but only some of them have RT requirements, the necessity for an I/O-Access API also on standard VMs is obvious. This issue is addressed by a further standards draft, the RTDA (RT Data Access) specification [4], produced by the RTDA working group of J-Consortium. The RTDA specification is incompatible to "RT Core " and RTSJ. Main focus of RTDA is the I/O-access mechanism, which is much more elaborate than in RTSJ and in RT Core. Functionally it corresponds to the approach of the Siemens product JFPC introduced in the following section.

## 3 OVERVIEW OF JFPC

JFPC is a Java class library (with OS-dependant modules in native code) providing an API for I/O data access, that is based on the ideas of OPC [5]. Data access to hardware registers is abstracted by proxy objects – the so called Items. Items are addressed by symbolic names, conforming to a URL-like notation, e.g.: */io/com202_1/slave_27/Bit/inbit_01*

This leads to a hierarchical structure of the namespace, reflecting the logical and physical architecture of the automation system, e.g. the existence of subordinate fieldbus systems

Access to JFPC is done via one central Server object, and the Items are subordinate objects. Items can be collected in groups, which contain also schedulable code (activation routine). Activation of groups can be done by timers, by interrupts or simply by monitoring the change of Items. All input Items of a group are read implicitly before calling the activation routine and all output Items are written implicitly after termination of the activation routine, thus implementing a PLC-like behaviour.

JFPC has a modular design, where the Server relies on a name service to locate Items and on providers to access the real hardware. There is a documented provider interface, enabling the extension of JFPC by new providers for new hardware components. Typically, providers consist of a Java part and a native code part, required to access the real hardware.

One specific provider - the so called "Remote Provider" - implements a proprietary application protocol above TCP, that allows client/server configurations of JFPC. So clients can access Items in the server via the Internet.

Each JFPC-System is configured dynamically during run time by reading a configuration file defining all target Items and the required providers with their associated parameters.

# 4 IMPLEMENTATION OP JFPC UNDER LINUX

## 4.1 General Approach

Forschungszentrum Jülich has received the complete source code of JFPC from Siemens. Main task of the port to Linux is the implementation of the so called RTS (run time system) provider. It is responsible for the handling of group activation. Thus it implements the timer services and the implicit pre-activation read of Itmes and post-activation write of Items, by calling optimized native C functions for reading and writing of items, which are registered at the RTS-Provider. Additionally specific providers for dedicated hardware have to be implemented. This includes Java code as well as native C code. Additionally Linux device drivers for all boards had to be implemented. We selected one digital output and and one digital input board as well as two PROFIBUS DP controllers.

PROFIBUS DP has been selected, because of its dominating market position. Thus a wide range of industrial equipment can be connected directly to this fieldbus and Forschungszentrum Jülich uses it as the common integral path to the world of industrial automation [6]. Hence evaluation of JFPC in the context of slow control applications requires the implementation of JFPC providers for PROFIBUS DP.

## 4.2 RTS-Provider

Only the native C code component of the RTS-Provider has to be changed in order to port it to a different operating system. The most important function to be implemented is *Java_DE_siemens_ad_jpc_server_RtsTimerEvent_waitTimerEvent0()*, which starts a timer and waits for its expiration in a thread context. This is simply mapped to the *nanosleep()* call, with additional use of *gettimeofday()* in order to reduce the jitter in the case of cyclic timers. Of course, all the timer control structures had to be modified in order to include process id and other additional bookkeeping data. An alternative approach using Linux interval timers with *setitimer()* has not been followed because of the unclear semantics of signals in a multithreaded environment.

## 4.3 Provider for the SMP16-EA216

SMP16-EA216 is digital output board by Siemens for the SMP (Siemens Microprocessor Periphery) bus. It has 32 optoisolated digital outputs organized in 4 independant ports. Signaling on the SMP bus is similar to the ISA bus, and the output ports are mapped by static configuration into the I/O space of the CPU. A Linux device driver has been implemented, which offers dedicated *ioctl()* calls, that do 8bit and 16bit write operations on these ports. Outputs are cached internally by the EA216 and can be read back. An additional bitmask parameter allows output of dedicated bits. The I/O base address of the board is an installation parameter of the device driver, and the device special name contains consecutive board numbers generated during installation (ea216_1 for board 1, etc). Implementation of the JFPC provider was straight forward, just by mapping the *writeItem* method to the appropriate *ioctl()* call. The URL-like name of an Item contains the special device file name as well as the byte or bit address in the range 0.0 to 3.7 (e.g. ..../ea216_1/Bit2.1 or .. /ea261_2/Byte0).

## 4.4 Provider for the SMP16-EA217

SMP16-EA217 is digital input board by Siemens for the SMP (Siemens Microprocessor Periphery) bus. It has 32 optoisolated digital inputs organized in 4 independant ports. The input ports are mapped by static configuration into the I/O space of the CPU. A Linux device driver has been implemented, which offers dedicated *ioctl()* calls, that do 8bit and 16bit read operations on these ports. Implementation of the JFPC provider was analogous to the EA216, just by mapping the *readItem* method to the appropriate *ioctl()* call.

## 4.5 Providers for the CPCI-FZJ-DP

CPCI-FZJ-DP is PROFIBUS DP controller in the CompactPCI formfactor developed by Forschungszentrum Jülich [6]. It supports only the cyclic services of PROFIBUS DP. PROFIBUS DP provides uses a simple programming model, because process data of slaves is mapped to a DPRAM via cyclic transfers on the PROFIBUS. An application on the master just reads and writes this DPRAM, asynchronously to the operation on the bus. The position of the process image of each slave in the DPRAM is determined statically by bus configuration. So basically a generic PROFIBUS DP provider has nothing else to do than to read and to write data in the DPRAM, by specifying the following parameters: PROFIBUS address of the slave, the length of the data, the relative start address of the data in the process image of the individual slave and a flag indicating input or output. An additional issue for JFPC is the interpretation of the data, because Java is strictly typed, and the representation of the above mentioned parameters in the JFPC configuration file

A possible approach for the implementation of providers would follow a hierarchical structure, with one provider for PROFIBUS DP, one provider for each type of slave, and additional providers for submodules in the case of modular slaves. We didn't choose this approach, because different slaves, even modular slaves, don't differ in their access mechanism, but only in the interpretation of data. So we chose a flat approach with different providers according to the interpretation of data and implemented the following four providers:

- Bit-Provider: generic provider for digital I/Os
- Word-Provider: generic provider for analog I/Os
- Provider for the Siemens stepper motor controller 1STEP
- Provider for Heidenhain encoders

All providers share a common native code part for accessing process data in the DPRAM. The Java code is device-specific, e.g, does scaling and format conversion in the case of the Word-Provider.

## 4.6 Provider for the CPCI-COM202

CPCI-COM202 is a PROFIBUS DP controller for CompactPCI, developed by Siemens. It supports not only the cyclic Services of DP but also DPV1 and DPV2 (isochronous PROFIBUS). The existing Windows NT driver of Siemens basically gives access to the DPRAM of the board. We implemented a device driver for Linux, that maps this functionality to appropriate *ioctl()* calls. Above the driver we implemented a library supporting the cyclic services of PROFIBUS DP, which was almost identical to the corresponding library for CPCI-FZJ-DP. Thus the corresponding Provider could be reused.

## FUTURE WORK

The JFPC port to Linux is still not completely stable. As soon as this problem is solved, performance analysis of the final system will be done. Also the jitter in group activations controlled by a interval timer will be analyzed. For the examination of RT behaviour, the implementation of JFPC under LynxOS, using HPs ChaiVM, is planned.